\documentclass[aps,twocolumn,superscriptaddress]{revtex4-1}
\usepackage{amsmath,amssymb}
\usepackage{graphics,graphicx}
\usepackage{dcolumn,bm}
\usepackage{psfrag}
\usepackage{color}
\usepackage{multirow}
\newcommand{\cu}
{\affiliation{Department of Physics, University of Calcutta,
92 Acharya Prafulla Chandra Road, Kolkata 700009, India.}}

\newcommand{\imsc}
{\affiliation{Institute for Mathematical Sciences, CIT Campus, Taramani, Chennai 600113, India.}}

\newcommand{\hbni}
{\affiliation{Homi Bhabha National Institute, Training School Complex, Anushakti Nagar, Mumbai 400094, India.}}

\begin{document}

\title{
Tagged particle dynamics in one dimensional $A+ A \to kA$ models with the particles biased to diffuse towards their nearest neighbour}

\author{Reshmi Roy}
\cu
\author{Purusattam Ray}%
\imsc
\hbni
\author{Parongama Sen}%
\cu

\begin{abstract}
Dynamical features of tagged particles are studied in a one dimensional $A+A \rightarrow kA$ system for $k=0$ and 1, where the particles $A$ have a bias $\epsilon$ $(0 \leq \epsilon \leq 0.5)$ to hop one step in the direction of their nearest 
neighboring particle. $\epsilon=0$ represents purely diffusive motion and $\epsilon=0.5$ represents purely deterministic motion of the particles. We show that for any $\epsilon$, there is a time scale $t^*$ which demarcates the dynamics of the particles. Below $t^*$, the dynamics are governed by the annihilation of 
the particles, and the particle motions are highly correlated, while for $t \gg t^*$, the particles move 
as independent biased walkers. $t^*$ diverges as $(\epsilon_c-\epsilon)^{-\gamma}$, where $\gamma=1$ and $\epsilon_c =0.5$. $\epsilon_c$ is a critical point of the dynamics. At $\epsilon_c$, the probability 
$S(t)$, that a walker changes direction of its path at time $t$, decays as $S(t) \sim t^{-1}$ and the 
distribution $D(\tau)$ of the time interval $\tau$ between consecutive changes in the direction of 
a typical walker decays with a power law as $D(\tau) \sim \tau^{-2}$.

\end{abstract}

\maketitle

\section{Introduction}

 Reaction diffusion systems in their simplest form with diffusion and annihilation of particles 
have been studied over the years \cite{privman, ligget, krapivsky,odor}. These are nonequilibrium systems
 of diffusing particles undergoing 
certain reactions. Depending on the nature of 
the problem, the particles could be molecules, 
chemical or biological entities, opinions in societies or market commodities. 
Such systems are frequently used to describe various aspects of wide varieties of chemical, 
biological and physical problems. In the lattice version of the single species problem, the 
lattice is filled with particles (say $A$) with some probability initially and at each time step, the particles 
are allowed to jump to 
one of the nearest neighbouring sites (diffusion) with a certain probability. The simplest form of 
particle reaction is when a certain 
number $l$ of the particles meet: $lA \rightarrow kA$ with $k<l$. It is well known that 
 annihilating random walkers with $l=2$  and $k=0$ corresponds to the 
Ising-Glauber kinetics while  the coalescing case with 
$l=2$ and $k=1$ describes the dynamics of the  
 $q$ state Potts model with 
$q \to \infty$, both at zero temperature and in one dimension \cite{derrida_95}.
Such systems have been studied in one
dimension \cite{racz,amar,avraham,alcaraz,krebs,santos,schutz,oliveira} as well as in higher dimensions 
\cite{kang,peliti,zumofen,droz}. 
Depending on the initial condition, whether one starts with even
or odd number of particles, the steady state will contain no particles or one particle respectively. 
The focus in all these analysis is how the system approaches the steady
state. In particular, one intends to know how the number of particles decays with time and the 
distribution of the intervals between the particles evolves with time.

Various reaction diffusion systems have been studied with different values of $k$ and $l$ in the past for 
different dynamical processes like ballistic annihilation\cite{krap_ballistic,bennaim,krap_ballisticanni_2001}, Levy walks \cite{albano_levy,krap_bennaim2016} and of course simple diffusion. However,
what happens if the dynamical process is intrinsically stochastic and diffusive is an important question which 
has not been studied much. The idea behind all these studies  is to find any universal behaviour in these system and the key factors which determine the universality. Here we ask the same question by introducing a bias which does not alter the 
existing features like conservation, range and nature of the interaction or the diffusional dynamics in the model.

We have studied  the model   $A+A\to kA$ where the particle $A$ diffuses with a preference towards its nearest neighbour.
Both the annihilating case ($k=0$) and the coalescing case ($k=1$) have been considered. It is important to note that this bias does not affect the annihilation process and  retains the 
Markovian property of the dynamics.  This simple extension, indeed, leads to  drastic changes in the 
bulk dynamical features. For $k=0$, the fraction of walkers $\rho(t)$ at time $t$ was found to decay as 
$\rho(t) \sim t^{-\alpha}$, where $\alpha \approx 1$ when the bias, however small, is introduced  \cite{soham_ray_ps2011, ray_ps2015}. In the absence of 
the bias, it is known that $\alpha = 1/2$. The value of $\alpha$ suggests that in the presence of the bias, the walkers, 
in the long time limit, behave as ballistic walkers.For the coalescing case with bias, the bulk
behaviour is identical, i.e.,   
$\alpha \approx  1$ (reported in the present paper). 

	The model considered in \cite{soham_ray_ps2011,ray_ps2015}  may thus appear to be equivalent to the system of annihilating ballistic walkers
at large times. But several features (e.g., persistence, domain growth etc.) of 
the dynamics show that it is actually not the same.
Hence, to get a better understanding we study the dynamics of a tracer walker in the biased case for both ($k=0$ and $1$), specifically to check  whether they perform ballistic motion or not.

In the following we briefly introduce the models and mention the different features studied and also the main results obtained.  We have a bunch of walkers on a one dimensional ring. At every time step, the walker hops 
one step to its left or right with a bias $\epsilon$ to move  in the direction of its nearest neighbour. 
$\epsilon =0$ implies no bias so that the walkers are purely random walkers 
and $\epsilon=\epsilon_c = 0.5$ implies full bias so that the walker always moves towards its nearest neighbour. 
Except for this point, the motion  is always stochastic.

For the annihilating case with $k=0$, we have a more detailed presentation
of the results. 
First, the probability $P(x,t)$ that a particle is at a distance $x$ from its origin after
time $t$ is estimated. We then calculate the probability that a walker changes its direction
as a function of time. The distribution of the time intervals over which
the walk continues in the same direction is also obtained. A change in the direction of motion can occur either due to diffusion or annihilation of the nearest particle(s).
We find that the dynamics of the walkers are controlled by two time regimes.
For time $t < t^*$, the dynamics are controlled by the annihilation of the particles. The motion  of the 
walkers, in this regime, is  highly correlated and the process is critical in the sense that there is no 
time scale in it. As a result,  the probability $S(t)$ of the change in the direction 
of the motion of the walker at time  $t$ decays with a power law; $S(t) \sim 1/t$. Similarly, the 
distribution $D(\tau)$ of the time interval $\tau$ spent between two changes in the direction of the motion 
of the walkers is scale free as $D(\tau) \sim 1/\tau^2$. We have found the full scaling behavior and arguments for 
the values of the exponents.  The crossover time $t^* \sim (\epsilon_c - \epsilon)^{-1}$, so  
$\epsilon_c$ can be interpreted as  a dynamical critical point  where a diverging time scale exists.

We have
also studied the coalescing model ($k=1$) with similar bias, i.e., $A + A \to A$ model.  
Without the bias, it is  equivalent to the $A + A \to \emptyset$ model as far as the decay of particles
in time is concerned.  In presence of bias, the scaling of the fraction of surviving particles  $\rho(t) \propto t^{-1}$ (details in section \ref{AAA}) shows that it is similar to  the annihilating model. 
The dynamics of the particles are indeed different in the coalescing model  as the distances  between the particles 
are  not much affected by a reaction,   except for the surviving particle that remains after the reaction.  
Here we have focussed on the behaviour of $S(t)$ and $D(\tau)$ and 
 find that the qualitative features of the dynamics of the tagged particle are again the same  as in the $A+A \to \emptyset$ model. 
However, here the crossover to the diffusion behaviour occurs at later times, so that $t^*$ is  higher in the $A+A \to A$ model. This is consistent with our inference that the early time regime is annihilation dominated as for $k=1$, the annihilation continues for a longer time.

\section{The  Model, dynamics and simulation details} \label{model}

The model consists of walkers denoted by $A$, undergoing the reaction 
$A+A \to kA$.
At each update, a site is selected randomly and if there is a particle on it, it moves towards
its nearest neighbour with probability $\epsilon+0.5$ $(0\leq\epsilon\leq0.5)$ 
and otherwise in the opposite direction. For $k=0$, if there is already 
another walker located on this neighbouring site, then both particles are annihilated and for $k=1$,
one of them survives.
Suppose, a walker is at site $i$ and its nearest neighbours are at
$i+x$ and $i-y$ on its right and left respectively; the walker will hop one step towards right with probability 
$0.5+\epsilon$ and to left with probability $0.5-\epsilon$ if $x<y$. 
In the rare 
 cases where the two neighbours are equidistant, the walker moves 
in either direction with equal probability. 

When the bias $\epsilon = 0.5$, the $k=0$ case corresponds to the spin model introduced in \cite{soham_ps2009}
(see Appendix for details).
 Hence, the dynamical updating scheme used here for $k=0$ has a one to one correspondence with the original spin dynamics used in \cite{soham_ps2009}.
 As the spin system in \cite{soham_ps2009} was considered to be highly disordered initially, we start with
a high density of walkers in this problem; specifically the number of walkers is chosen to be $L/2$
on a one dimensional lattice of size $L$.
	To maintain the correspondence with spin dynamics,  
the walkers are updated asynchronously and at each update a site is chosen randomly, rather than a walker, for updating. 
One Monte Carlo step (MCS) comprises of $L$ such updates. The same dynamical scheme was used
 in 
\cite{ray_ps2015,soham_ray_ps2011} where the bulk properties of the walker model were studied. 

 The dynamical scheme allows the possibility that a walker's state
may not be updated at all. This is because if a site is not selected, the position of the corresponding 
walker will not be updated. 
 It may also happen that a walker is updated more than once
in the following manner: if a walker moves to the site $j$ and the site $j$ is selected later, 
then the position of the same walker is updated again.
This signifies that the net displacement of a particular walker may even be zero when it performs more than
one movement in the same MCS.
For all calculations, the final positions of the walkers after the completion 
of one MCS are considered. 
The results reported here are for simulations done on  lattices of size 12000 or more and the maximum number of configuration studied was 2000. Periodic
boundary condition has been used for all the simulations.

In the $A + A \to A$ model ($k=1$), the same dynamical scheme is used. Here, once two walkers meet, one of them
will survive. In order to study the tagged particle dynamics, we need to label the surviving particle. We use  the convention that the particle which makes the last movement survives. We have checked that the random  convention  (either
of the two particles is taken to be the survivor randomly) leads to the same results
qualitatively.  
 
\section{Results  for $A+A \to \emptyset$ $(k=0)$ model}
\label{AA0}

\begin{figure}
\begin{center}
\includegraphics[width=9cm]{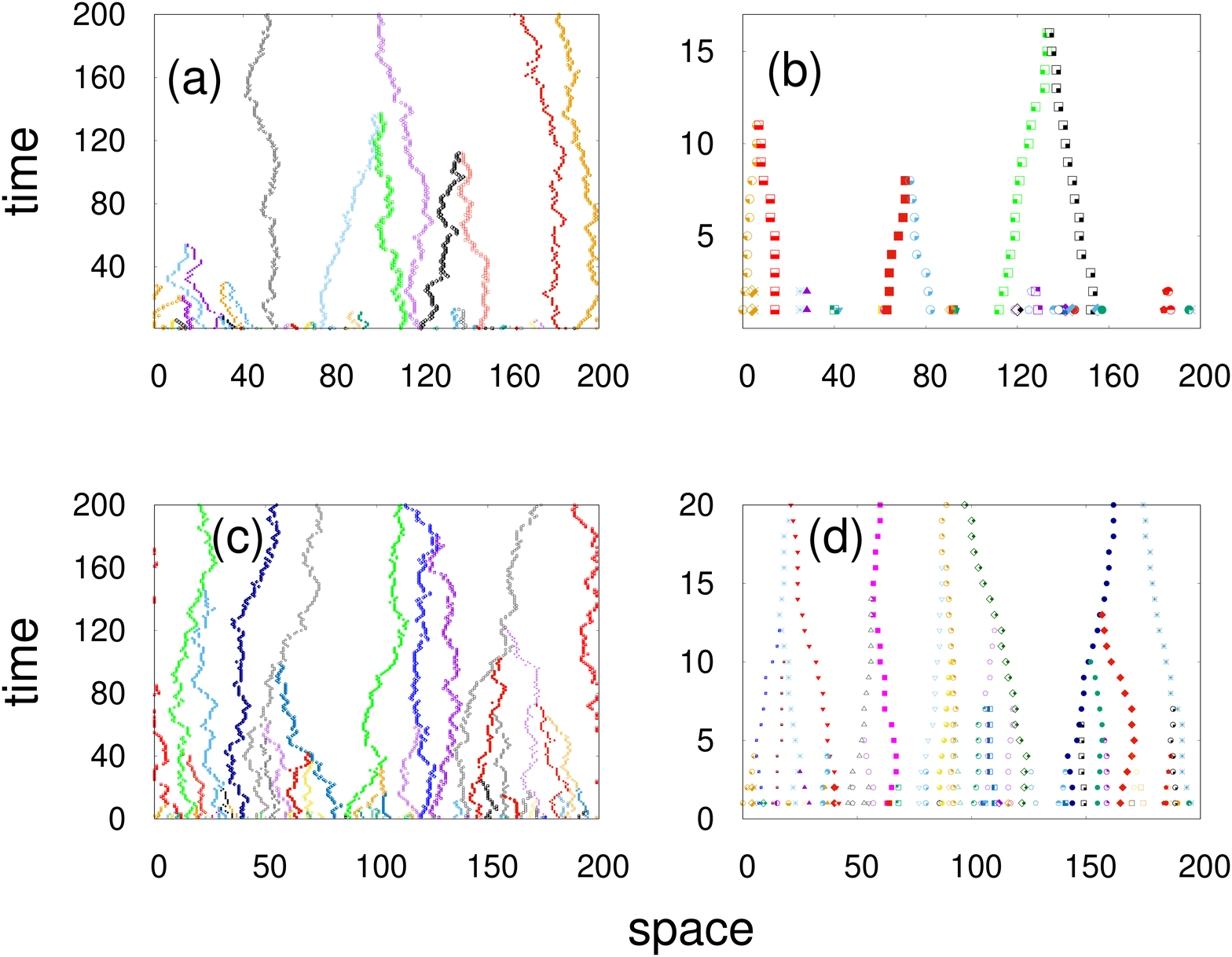}
\end{center}
\caption{Snapshots of the system at different times for  $\epsilon$=0 (a) and  $\epsilon$=0.5 (b) for $A+A \to \emptyset$ model.
Lower panel show the snapshots for $A+A \to A $ model for  $\epsilon$=0 (c) and  $\epsilon$=0.5 (d).
 The trajectories in different colors represent different particles.}
\label{snap_asyn}
\end{figure}

To check the movement of individual walkers we took snapshots of the system 
at different times. Fig. \ref{snap_asyn}(a) and (b)  show the world lines of the motion of the particles
for $\epsilon=0$ and $\epsilon=0.5$. It clearly shows that the motion of the individual particles in 
the two extreme cases are remarkably different. Annihilation dominates for $\epsilon=0.5$ while for $\epsilon=0$ the walk is diffusive as expected. For the intermediate values of $\epsilon$, both the mechanisms of diffusion and annihilation will be important and thus,
as we will see later, give rise to the crossover effect for the system. To probe the 
dynamics of the particles,
we have studied the following three quantities: 
(i) the probability distribution $P(x,t)$ of finding a particle $A$ at distance  $x$ from its origin at time $t$, (ii) the 
probability $S(t)$ of the change in the direction in the motion of a walker at a time $t$ 
and (iii) the distribution $D(\tau)$ of the time interval $\tau$ between two successive changes in the 
direction of the motion of a walker. The results for each of these quantities are described in the 
following three subsections.

\subsection{Probability distribution $P(x,t)$}

For $\epsilon=0$, the single particle motion is diffusive and the corresponding probability distribution 
$P(x,t)$ is known to be Gaussian. This remains true even in the presence of annihilation.

\begin{figure}
\includegraphics[width=9cm]{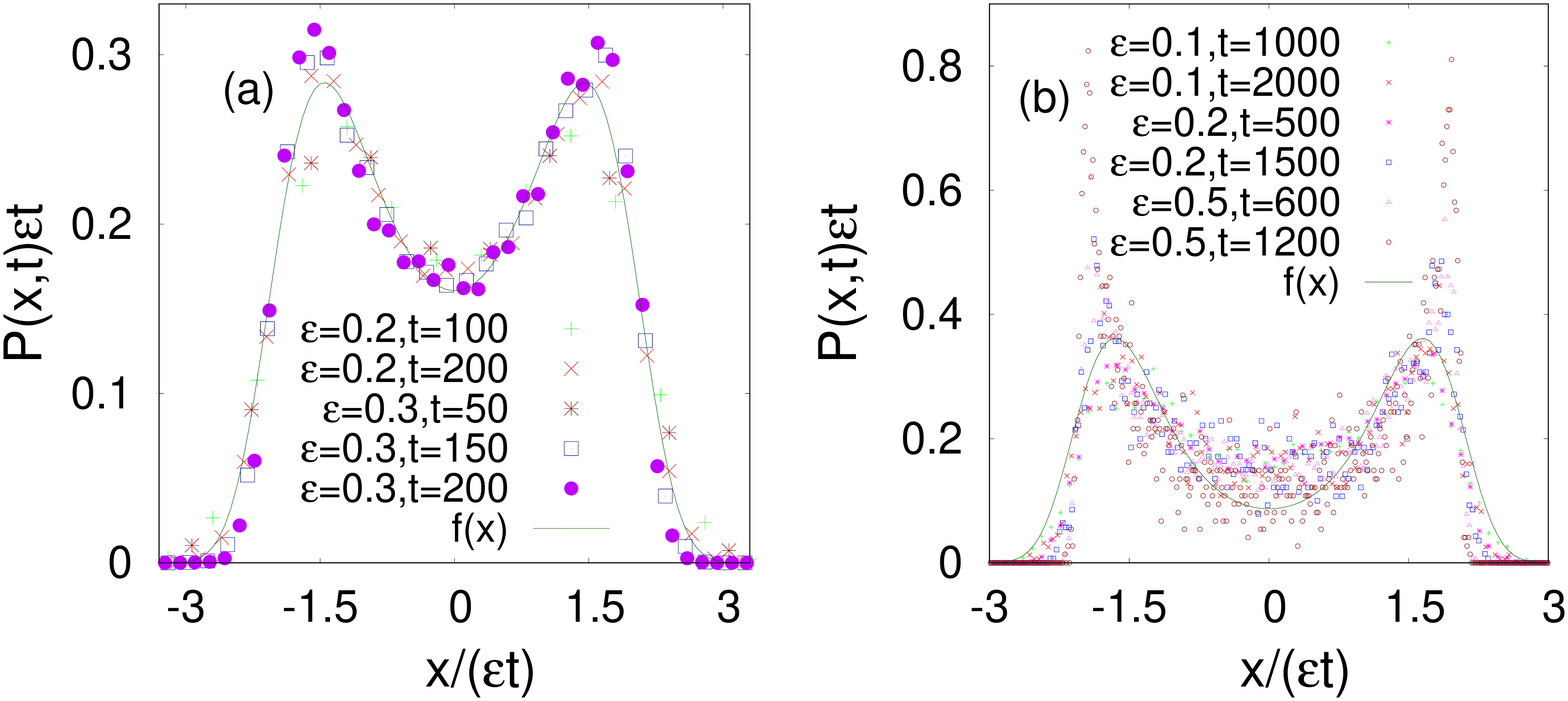}

\caption{(a): Data collapse of ${P(x,t)\epsilon t}$ against ${x/\epsilon t}$ for $\epsilon=0.2$ and $0.3$ are shown at early time regime, (b): Data collapse of ${\it P(x,t)\epsilon t}$ against ${ \it x/\epsilon t}$ are shown for $\epsilon=0.1,0.2,0.5$ at late time regime. These data are for the $A+A \to \emptyset$ model.}
\label{probdistri}
\end{figure}

For $\epsilon \neq 0$, $P(x,t)$ changes drastically. 
The distributions are still symmetric as the motion
 of individual particles can occur in both directions (left and right).
However, there is no peak at the origin ($x=0$) and 
instead a double peak structure emerges 
with a dip at $x=0$.
To obtain a collapse of the data at different times, we note that the scaling variable is $x/\epsilon t$ for all values of $\epsilon$.
We find that  the collapsed data can be fit to the form 

\begin{equation}
 P(x,t)\epsilon t  = 
f(\frac{x}{\epsilon t}) 
\propto \exp[-\beta\{(\frac{\gamma x}{\epsilon t})^2-1\}^2].
 \label{distri_function}
\end{equation}
The data collapse in the early time regime is shown in Fig. \ref{probdistri}a.
However, the data collapse as well as the above scaling form seems to be less accurate at later times. 
On investigating further, we find that while we attempt to fit the data
individually for each $\epsilon$ and $t$ by  the  form given in Eq. (\ref{distri_function}), 
only in the early regime ($\epsilon t \lesssim  100$),
both $\beta$ and $\gamma$ are constants. 
While  $\gamma$ shows negligible dependence on $\epsilon$ and $t$,
$\beta$ strongly depends on $\epsilon t$;  beyond 
$\epsilon t =  100$ it is no longer a constant but increases sharply as
a function of $\epsilon t$. 
Hence the distribution
scaled in the above manner shows a dip at the center which goes down  with time while  the 
peak heights increase such that the   data do not collapse well as  shown in Fig. \ref{probdistri}b.

The above study suggests that  at vary late stages, the scaled distribution 
 will assume a double delta functional form and a universal scaling function 
exists only in the early time regime ($\epsilon t \lesssim  100$). 
We can relate the breakdown of the universal behaviour to the
crossover phenomena that is revealed more clearly in the following subsections.

\subsection{Probability of change in direction} \label{prob_dirchange}

The probability of direction change at time $t$ is obtained by estimating 
the fraction of walkers that change direction at time t. For $\epsilon=0$, as the system is diffusive, the
probability of direction change $S(t) = p_0$, a constant
independent of time. For a purely diffusive random walk, $p_0= 0.5$. 
But here asynchronous dynamics have been used and this updating scheme
allows the walkers to remain 
in the same state within a MCS as already discussed in the previous section. 
This dynamics can only decrease the probability of change in direction. 
$p_0$ for $\epsilon=0$ actually turns out to be $\approx 0.27$ numerically.

\begin{figure}
\begin{center}
\includegraphics[width=8cm]{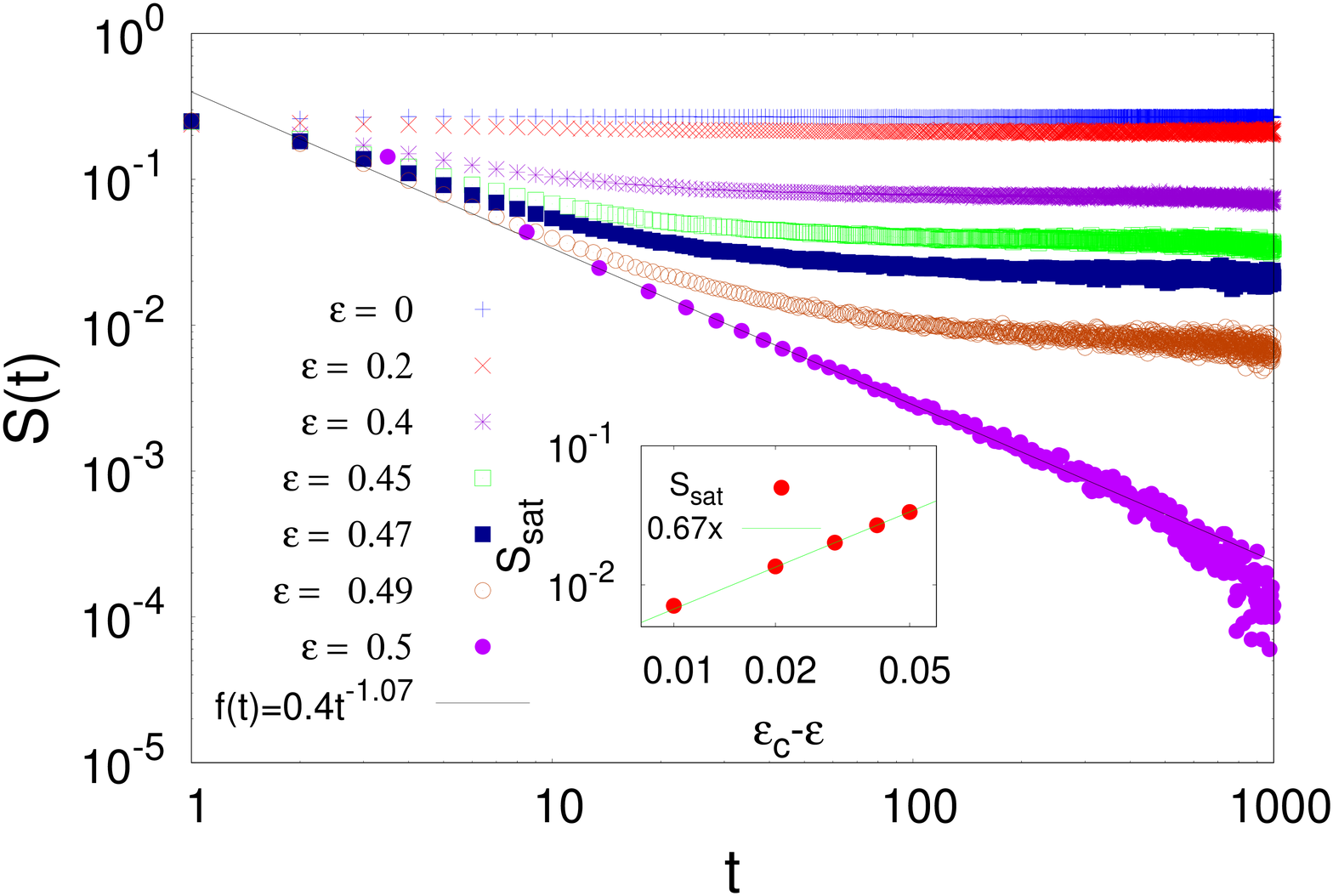}
\end{center}
\caption{Probability of direction change of tagged particle for different $\epsilon$ in the $A + A \to \emptyset$ model. For $\epsilon=0.5$, 
it decays as $t^{-1}$. Inset shows the variation of $S_{sat}$ with $\epsilon_c-\epsilon$. }
\label{dir_change}
\end{figure}

For $0<\epsilon<0.5$, the change in direction of a walker occurs 
due to two reasons; either due to the annihilation of a neighbouring walker or
because of the diffusive component which is large for small $\epsilon$. At earlier times, 
the walker density is large and so the number of annihilation is considerable. Therefore the change in 
direction of the walkers is dominated by the annihilation process. However, as time progresses, 
 annihilation becomes rarer and therefore the diffusive component 
becomes the dominating factor. So a saturation value $S_{sat}$ of $S(t)$ is reached at 
a later time, typically after a time $t^*$. The data for $S(t)$ is shown in Fig. \ref{dir_change} and 
the inset shows the variation of $S_{sat}$ with ${\epsilon_c-\epsilon}$ where $\epsilon_c = 0.5$. 
As expected, $S_{sat}$ decreases as $\epsilon$ is made larger. 
In fact, we find that unless $\epsilon$ is very close $\epsilon_c$, the saturation 
is reached very fast, typically within one hundred MC step. $S_{sat}$ shows a
 linear variation with $\epsilon_c-\epsilon$, shown in inset of Fig. \ref {dir_change}.

\begin{figure}
\includegraphics[width=9.6cm]{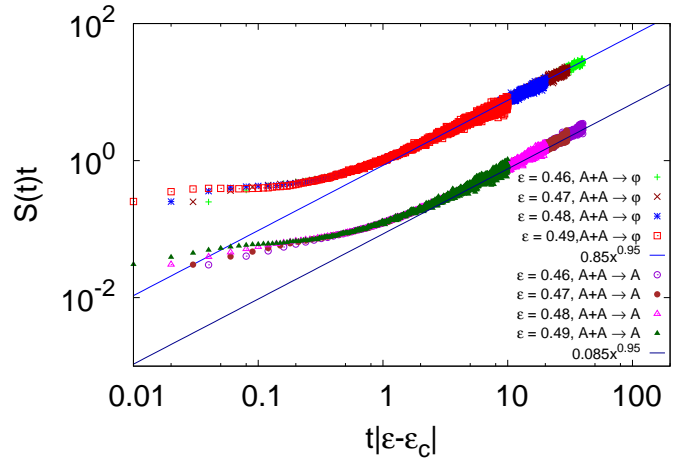} 
\caption{Variation of $S(t)t$ with $t|\epsilon-\epsilon _c|$ shows a data collapse for both the models $A + A \to \emptyset$ and $A + A \to A$, where $\epsilon _c=0.5$.
Data for the $A + A \to A$ model have been shifted along $y$ axis for better clarity. The linear regions in the log-log plot are fitted to power law forms with the exponent very close to unity.}
\label{scale_dir}
\end{figure}

One can obtain a data collapse by plotting $S(t)t$ against $t(\epsilon _c - \epsilon)$, shown in Fig. \ref{scale_dir}. 
This indicates that one can write $S(t)$ as
\begin{equation}
S(t)=\frac{1}{t} g(z),
\label{dirscale}
\end{equation}
where $z=t(\epsilon_c  - \epsilon)$ and $g(z)$ is a scaling function. $g(z)$ is constant for $z<1$ and $g(z) \sim z$ 
for large $z$. Therefore, $p_\epsilon \equiv S(t \to \infty) \propto (\epsilon_c - \epsilon)$, 
which is consistent with the variation of $S_{sat}$ with $(\epsilon_c - \epsilon)$ (see inset of Fig. \ref{dir_change}).
Hence one can argue that $t^* = (\epsilon _c - \epsilon)^{-1}$ 
acts as a timescale, below which $S(t) \propto t^{-\delta}$ with $\delta=1$. As $t^*$ diverges at $\epsilon_c$, 
there is no saturation region for $\epsilon=\epsilon_c$ and $S(t)$ shows a power law decay, $S(t) \sim t^{-1}$ for all times as
shown in Fig. \ref{dir_change}. The divergence of $t^*$ as $\epsilon \rightarrow \epsilon_c$ justifies  that $\epsilon_c$
is the dynamical critical point.

One can argue that the value of $\delta$ is unity for the deterministic case 
$\epsilon = 0.5$, where the walker always 
moves towards its nearer neighbour. A direction change can occur only 
if an adjacent walker is annihilated (however, this is a 
 necessary but not sufficient condition). 
Let $A(t)$ be the number of annihilation taking place at time $t$.
If $N(t)$ is the number of walkers at time $t$, $A(t)$ is given by 
$- \frac {dN}{dt} \propto t^{-\alpha - 1} = t^{-2}$.       
Since $N(t)$ is proportional to $t^{-1}$ and $S(t)$ is proportional to 
$A(t)/N(t)$, therefore $S(t) \sim t^{-1}$. It may be added here that $A(t)$ and $N(t)$ have the same behaviour for all
$\epsilon \neq 0$, 
however, for $\epsilon \neq 0.5$, direction change may occur even when there is no annihilation. 
The above argument is valid only for $\epsilon = 0.5$ for which there is no diffusive component.
However, the fact that $S(t) \propto t^{-1}$ in the early time regime for $\epsilon\neq 0.5$
also, shows that the annihilation plays the key role in the dynamics here; 
the diffusive component is virtually ineffective.  Clearly a crossover behaviour 
occurs in time. 
The crossover occurs at a  time 
when annihilation becomes rare. This depends on two factors: the density 
of the walkers and the strength of the bias. In time, the density decreases and   
 beyond the crossover  time $t^*$, the bias is not strong enough to cause two particles to 
come close enough and cause an annihilation.  The  motion effectively becomes uncorrelated. Obviously, the crossover 
occurs at later times as $\epsilon$, representing the bias, becomes larger and the inherent diffusive component becomes 
weaker  making annihilations more probable. Therefore, at  $\epsilon = 0.5$, the fully biased point,
  $S(t) \propto t^{-1}$ 
and the crossover time diverges.

The nature of the walk remains ballistic in all regimes due to the bias,
however small, to move towards the nearest neighbours. 
This is consistent with the conjecture that the probability distribution assumes
a double delta form at large times mentioned in the last subsection. 

\subsection{Distribution of time intervals between 
consecutive change in direction}

\begin{figure}[h]
\includegraphics[width= 9cm]{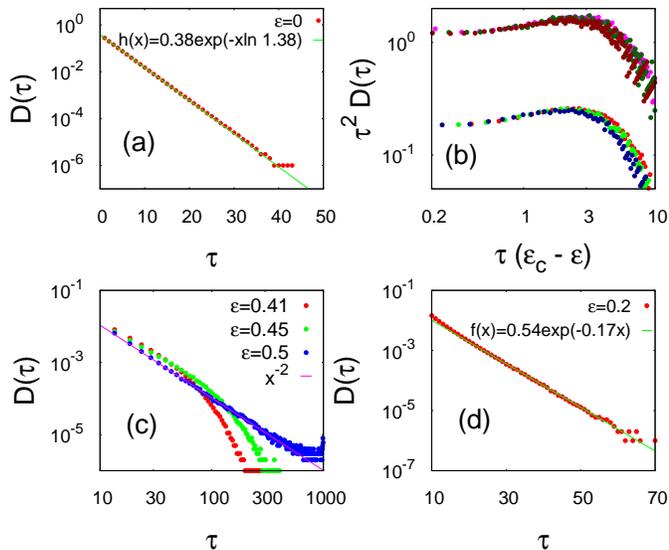}
\caption{(a), (c) and (d) show data for the $A + A \to \emptyset$ 
model while (b) includes the data for the $A + A \to A$ model also. 
(a)  Variation of $D(\tau)$ over $\tau$ for $\epsilon=0$
(b) Collapsed data for $D(\tau) \tau^2$ against $\tau (\epsilon_c - \epsilon)$ for $\epsilon=0.475,0.48,0.485$.
Data for the $A + A \to A$ model (lower curves) have been shifted for better clarity. (c) shows the data for $D(\tau)$ against $\tau$
for $\epsilon=0.41,0.45,0.5$, where power law decay exists over a small time interval 
and power law region decreases with the decrease of $\epsilon$. (d) shows the variation of $D(\tau)$ against $\tau$ for $\epsilon=0.2$ and the data is fitted according to Eq. (\ref{timedist_expo}) for $\tau \gg t^*$ for  $A+A \to \emptyset$ model.}
\label{timedist}
\end{figure}

Another interesting quantity is $D(\tau)$, the interval of time $\tau$ spent without change 
in direction of motion. For random walkers with $\epsilon=0$, the probability that in 
the time interval $\tau$, there is no direction change is given by

\begin{equation}
D(\tau)={p_0}^{2}({1-p_0})^{\tau}.
\end{equation}
This reduces to an exponential form: $D(\tau)\propto 
\exp[-\tau \ln \{1/(1-p_0)\}]$. Fig. \ref{timedist}a shows the data for $D(\tau)$ for $\epsilon=0$. From the numerical simulation, we find $D(\tau) \sim \exp(-\tau \ln 1.38)$ for $\epsilon=0$, which is consistent with $p_0 \approx 0.27$.

For general values of $\epsilon \neq 0$, we note that $D(\tau)$ 
obeys the following form 

\begin{equation}
D(\tau) \sim \frac{1}{\tau^2} \phi(z), 
\label{timedist_expo}
\end{equation}
where $z = \tau(\epsilon_c-\epsilon)$ is the scaling argument and 
$\phi(z)$ is the scaling function. 
$\phi(z)$ is constant for $ z<1$ and 
proportional to $\exp(-z)$ for $z \gg 1$. The data are shown in Fig. \ref{timedist}b.

Thus it is indicated that here also a crossover behaviour occurs at $\tau=\tau ^*$ 
with $\tau^* \propto (\epsilon_c -\epsilon)^{-1}$, beyond which the exponential decay is observed 
and below which there will be a power law behaviour. Obviously for $\epsilon=0.5$, $\tau^*$ diverges such that 
only the power law decay will be observed with an exponent 2 which is indeed the case as shown in Fig. \ref{timedist}c.

It can be argued why the exponent is $2$ for $\epsilon=0.5$. Suppose the walker moves without direction 
change in the interval $t_0+1$ to $t_0+\tau$. This means it changes direction at times $t_0$ and $t_0+\tau+1$. 
Hence, $D(t_0,\tau)$ is given by 
\begin{eqnarray}
D(t_0,\tau)&=&S(t_0) S(t_0+\tau+1) \prod _{x=1}^{\tau}[1-S(t_0+x)].
\nonumber
\end{eqnarray}
Using the variation of $S(t) \propto 1/t$ obtained in the last subsection, 
\begin{eqnarray}
D(t_0,\tau)& \propto &(t_0^{-1}) (t_0+\tau+1)^{-1} \prod _{x=1}^{\tau}(1-\frac{1}{t_0+x}).
\label{timeeq3} 
\end{eqnarray}

Taking logarithm of both sides of Eq. (\ref{timeeq3}) and converting summation into an integral, one gets

\begin{eqnarray}
\ln D(t_0,\tau)&=&-2 \ln(t_0+\tau), 
\nonumber
\end{eqnarray}
apart from a constant factor.
One can always choose the origin $t_0$ to be zero, such that 
\begin{eqnarray}
 D(\tau) \sim \tau^{-2}
\end{eqnarray}
showing consistency with the numerical results. 
(Fig. \ref{timedist}b). 

\begin{figure}[h]
\includegraphics[width=8cm]{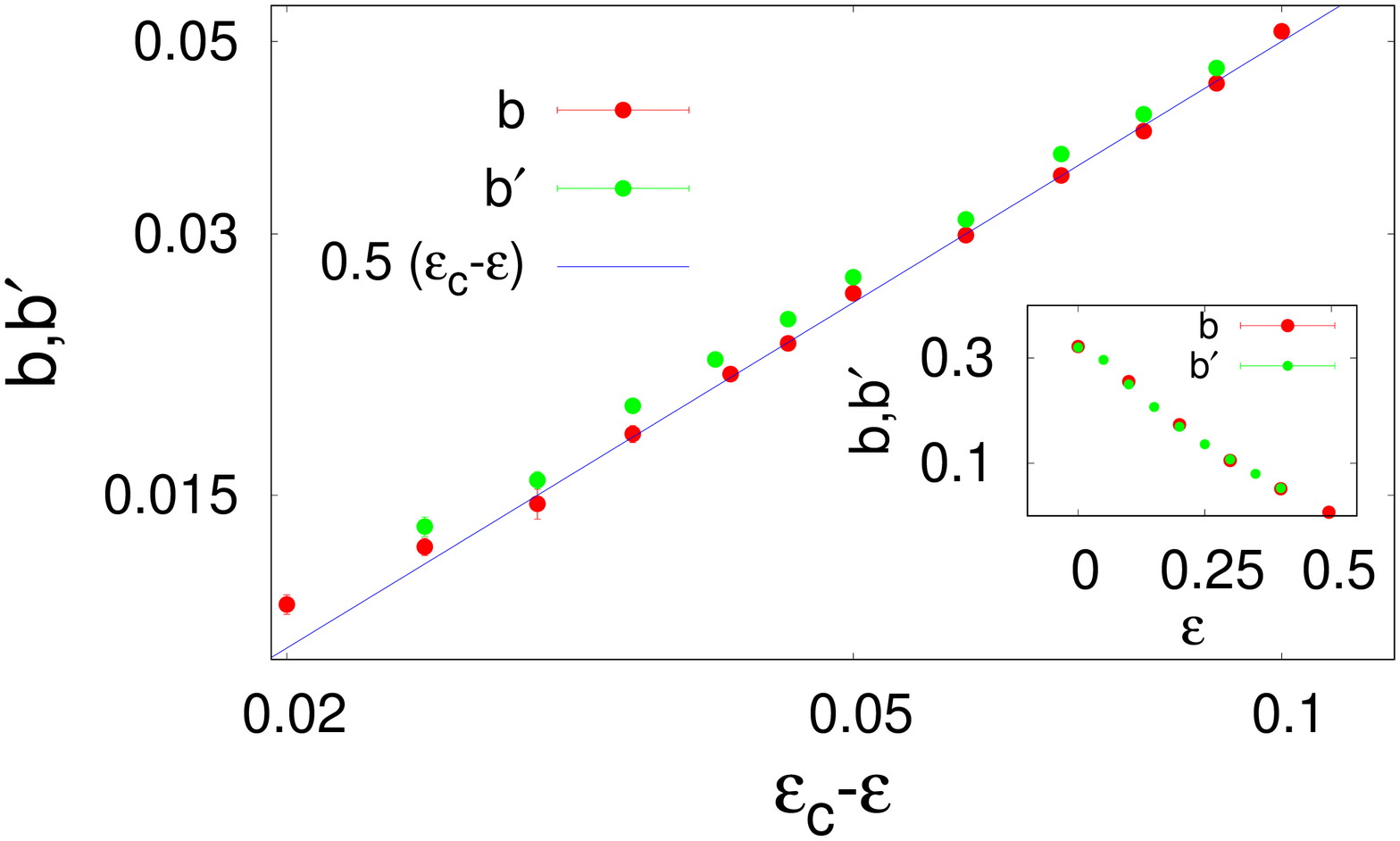}
\caption{Variation of $b$ and $b^\prime$ with $\epsilon_c-\epsilon$  shown in a log-log plot when $\epsilon$ is very close to 0.5
and inset shows the variation with $\epsilon$ for the full range.}
\label{bvariplot}
\end{figure}

One can also justify the crossover behaviour for $0<\epsilon<0.5$. Here, the crossover behaviour in $S(t)$ found in Sec. \ref{prob_dirchange}, should be taken into account while calculating $D(\tau)$.  
$S(t)$ decays in a power law manner at short times to a constant value in the late time regime. The  relatively larger value of $S(t)$
 will be responsible for the behaviour of $D(\tau)$ for small $\tau$. Hence, for
 small $\tau$, the power law behaviour of $S(t)$ will be relevant for which it has been already shown that $D(\tau) \propto \tau^{-2}$. On the other hand, the 
constant (lower) value of $S(t)$ will be responsible for contribution to $D(\tau)$ for
large values of $\tau$. For $t>t^*$,
$S(t) = p_{\epsilon} = a_0(\epsilon_c - \epsilon)$, where $a_0$ is a constant less than unity (see Fig. \ref{scale_dir}). Using this value, one gets therefore
$D(\tau)= (p_{\epsilon} )^2 (1-p_{\epsilon})^{\tau} \sim
\exp(\tau \ln (1- a_0(\epsilon_c - \epsilon)))$. As 
 $a_0(\epsilon_c - \epsilon)$ is less than unity, the expression for $D(\tau)$ 
simplifies to
\begin{eqnarray}
D(\tau) \sim \exp(-a_0(\epsilon_c - \epsilon)\tau).
\label{expfit}
\end{eqnarray}
$D(\tau)$ can indeed be fit to an exponential form for large values of $\tau$
(see Fig. \ref{timedist}d):
 $D(\tau) \sim \exp(-b\tau)$ 
(as long as $\epsilon$ is not very close to $\epsilon_c$ for reasons that will be clarified later) 
and $b$ can be fitted to the form 
\begin{equation}
b=b_0 (\epsilon_c-\epsilon), 
\label{cfit}
\end{equation}
where $b_0=0.5$, shown in Fig. \ref{bvariplot}.
This agrees with the expectation that $b$ should be varying linearly with
$(\epsilon _c -\epsilon)$ as indicated by Eq. (\ref{expfit}). 
It is also observed that $b$ approaches the value $\ln 1.38$ as $\epsilon \rightarrow 0$
and $b\rightarrow 0$ as $\epsilon \rightarrow \epsilon _c =0.5 $ (Fig. \ref{bvariplot}). 

Fig. \ref{timedist}b shows that for large values of the argument, beyond the crossover, 
the data collapse is not of very good quality. 
This is because as $\epsilon$ approaches
0.5 the crossover time increases and the exponential behaviour exists only for very large 
values of $\tau$ where the statistics is obviously poorer. 
This is the reason for which the estimation of $b$ for $\epsilon \rightarrow \epsilon_c$ 
becomes less reliable as mentioned before. On the other hand, to show the power law region 
one has to use values of $\epsilon$ fairly close to 0.5.

\section {Results for $A+ A \to A$  ($k=1$) model}
\label{AAA}

\begin{figure}[h]
\includegraphics[width=7cm]{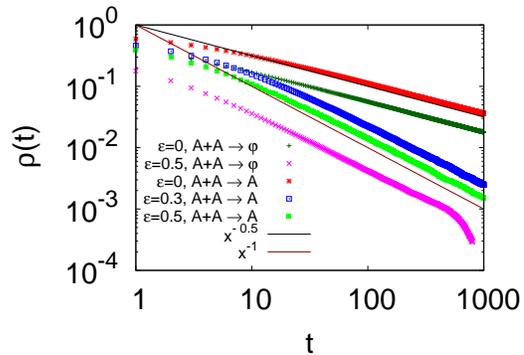}
\caption{Variation of $\rho(t)$ with $t$ is shown in a log-log plot for different values of $\epsilon$ for  $A+A \to \emptyset$ and $A+A \to A$  models. }
\label{alive_both}
\end{figure}

For the $A + A \to A$ model, the $\epsilon =0$ case is known to have the 
scaling form for the fraction of surviving particles as $\rho(t)
\propto  t^{-1/2}$ \cite{krap_ballistic}.   
In  the biased case, with  any $\epsilon \neq 0$ we find that the scaling is again like the $A + A \to 0$ case (with bias) as $\rho(t) \propto t^{-1}$ shown in Fig. \ref{alive_both}. 
Typical snapshots of the walk are shown in Fig. \ref{snap_asyn}(c) and (d).

For the motion of the  tagged particles in  the $A + A \to A$ model, we restrict the study to the probability of direction change and distribution of the time interval of motion executed without direction change. 
Again we find no significant change from the behaviour for the $A + A \to 0$ case., i.e., here also $S(t) \propto t^{-1}$ for $\epsilon = 0.5$ while for other values of $\epsilon$, there is a crossover to a diffusive behaviour. 
In fact, when  $S(t)t$ is plotted against $t(\epsilon_c -\epsilon)$, 
we again find that the scaling function has a constant part and a linear variation 
at larger values of the scaled variable (Fig. \ref{scale_dir}).

One can, in fact, use the same argument to justify the scaling behaviour $S(t) \propto t^{-1}$ for $\epsilon = 0.5$. This is because in this case also, the only way 
the direction change can take place is through annihilation. However, there is a subtle difference. For the $A + A \to \emptyset$ model, when two particles are annihilated, 
direction change can take place for their neighbouring particles. On the other hand, in the $A+A \to A $ case, the direction change may occur for the surviving particle while its neighbouring particles 
usually remain unaffected (see Fig. \ref{snap_asyn}). 
Another important point to note is that in the scaling function for $S(t)t$, the linear fitting is 
appropriate beyond a larger value of the scaled variable, i.e., the 
crossover to diffusive behaviour takes place later in the $A + A \to A$ model
in comparison (see Fig. \ref{scale_dir}). 
This is consistent with our inference  that the early time regime is annihilation dominated as 
the annihilations in the $A + A \to A$ continue for a longer time.

The distribution for the time intervals of motion without change in direction
again shows similar scaling. 
In Fig. \ref{timedist}(b), we show the comparative behaviour 
for the two models. The  tail of the  scaling function is  obtained
 once again  as  $\exp(-b^\prime \tau)$,  where  $b^\prime $ shows a linear variation with $(\epsilon_c - \epsilon)$ (Fig.  \ref{bvariplot}).

\section{DISCUSSIONS}

We have studied the motion of the tagged particles $A$, in one dimension, undergoing the reaction 
$A+A\rightarrow kA$ with $k=0$ and 1 with the additional feature that a particle walks with a probability 
$0.5+\epsilon$ towards its nearest neighbour and with a probability $0.5-\epsilon$ in the 
other direction. This is perhaps one of the simplest models which exhibits critical dynamics. 

The particles, when $\epsilon =0$, perform normal random walk, so their motions are not 
correlated. The reaction makes the fraction $\rho$ of particles decay with time $t$ as 
$N(t) \sim t^{-\alpha}$ with $\alpha=1/2$. For any non-zero $\epsilon$, the value of $\alpha$ has 
been found to be altered to 1. The value of $\alpha=1$ suggests that the particle motion is not 
random anymore but is ballistic. However, it has to be remembered that $A+A \rightarrow \emptyset$ 
model with ballistic walkers $A$ do not correspond to $\alpha=1$ and the results depend on the 
distribution of initial velocities of the particles \cite{krap_ballistic,bennaim}.
	
	Studying the tagged particles reveal that the effect of $\epsilon$ in conjunction with the annihilation 
reaction makes the dynamics of the particles correlated over a large time scale. This time scale depends 
on $\epsilon$ and diverges at $\epsilon=0.5$. Consequently, the dynamics 
become critical, in the sense that, the probability $S(t)$ of the particles to change the direction of their 
motions reduces with time as $1/t$ and the distribution $D(\tau)$ of time interval $\tau$ over 
which the particles on average move along the same direction follows power law: $D(\tau) \sim 1/\tau^2$.

Detailed study of $S(t)$ and $D(\tau)$ shows that there is a crossover from the annihilation dominated regime to a (partially) diffusive regime at time $t^* \propto (\epsilon_c-\epsilon)^{-1}$.
Beyond $t^*$, $S(t)$ is a constant for $0 < \epsilon < 0.5$, although the actual value is less compared to the 
unbiased case $\epsilon =0$. However, the overall motion is still ballistic, $\langle |x| \rangle \sim t$, for any $\epsilon> 0$
because of the presence of the bias. 
This is supported by the behaviour of the distribution $P(x,t)$ tending towards a double delta function 
(studied for the $k=0$ model)
at very late times while for $\epsilon=0$, the distribution is always Gaussian.

It may be mentioned here that 
the change in the behaviour of the probability distribution from a Gaussian for $\epsilon =0$ to a bimodal form for $\epsilon \neq 0$ is reminiscent of the order parameter distribution above and below the critical temperature for Ising like systems; the form in Eq. (\ref{distri_function}) is also similar to the case for continuous spins.

In conclusion, we have shown how the  bias to move towards nearest neighbours generates 
correlation in the motion of the  particles in a simple
$A+A \rightarrow kA$ reaction process. Also, we conclude that the divergences in the timescales  and power law behaviour in the relevant dynamical variables indicate that $\epsilon_c = 0.5$ is a dynamical critical point. In
the present study we have detected a crossover from a correlated to a individual motion scenario in the presence of the bias. Simultaneously we obtain two new dynamical exponents using Monte Carlo simulation and simple arguments and calculation. The reaction is not dependent on the bias and except for the point $\epsilon=0.5$, the motion is still stochastic. 
The present study is able to manifest at the individual level the precise 
role of the bias and how the dynamics are different from simple 
ballistic motion.

Acknowledgement: The authors thank DST-SERB project, File no. EMR/2016/005429 (Government of India) for financial support. Discussions with Soham Biswas is also acknowledged.

\section{appendix}

\begin{figure}[h]
\includegraphics[width=10cm]{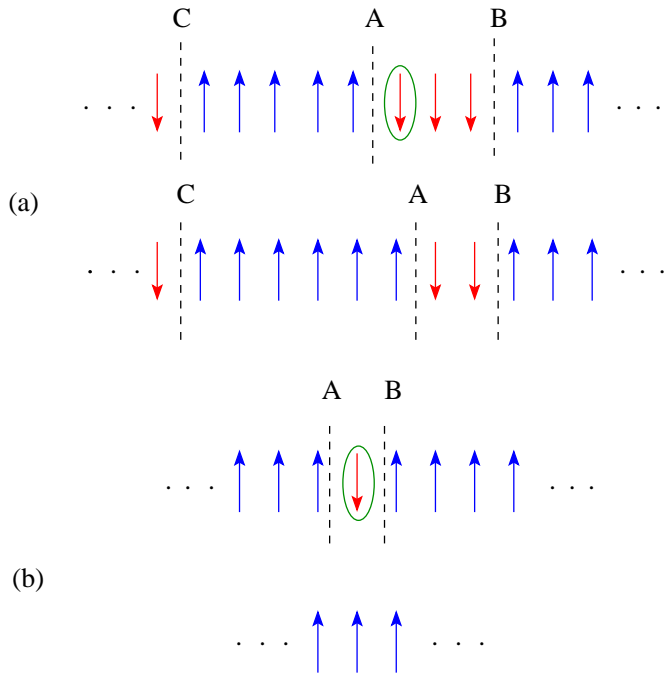}
\caption{A schematic picture of the dynamics taking place in the model proposed in \cite{soham_ps2009}. 
(a) Case I: Here, the highlighted spin changes its state as the neighbouring domain of down spins
 is of size two while the size 
of the other neighbouring domain of up spins is five. 
 Equivalently, in the walker picture, 
the interface $A$ moves towards $B$, which is closer to it compared to $C$. \\
(b) Case II: When a down (up) spin is sandwiched between up (down) spins, it will always flip which leads to annihilation of $A$ and $B$.}

\label{spin_fig}
\end{figure}

Here we argue that the spin model proposed in \cite{soham_ps2009} has a one to one correspondence with the particle/walker model when $\epsilon = 0.5$ for $k=0$. 
In \cite{soham_ps2009}, spins with state $\pm 1$ are considered on a one dimensional lattice. A spin flips when it sits at the boundary of two domains of oppositely oriented spins. At subsequent times, the state of the spins is determined by the size of the two neighbouring domains; it is simply changed to the sign of the spins in the larger domain. Thus the smaller domain shrinks  further and one can have an equivalent picture of a particle which moves towards its nearest neighbour. The scheme is illustrated in Fig. \ref{spin_fig}.

\begin{thebibliography}{99}
\bibitem{privman} Privman V., ed. {\em Nonequilibrium Statistical Mechanics in One Dimension}, 
Cambridge University Press, Cambridge (1997).
\bibitem{ligget} Ligget T. M., {\em Interacting Particle Systems}, Springer-Verlag, New York, (1985).
\bibitem{krapivsky} Krapivsky P. L., Redner S.  and Ben-Naim E., {\em A Kinetic View of Statistical 
Physics}, Cambridge University Press, Cambridge (2009). 
\bibitem{odor} Odor G., Rev. Mod. Phys. {\bf 76}, 663 (2004).
\bibitem{derrida_95} Derrida B., J. Phys. A Math. Gen. {\bf 28}, 1481 (1995).
\bibitem{racz} Racz Z., Phys. Rev. lett. {\bf 55}, 1707 (1985).
\bibitem{amar} Amar J. G. and Family F., Phys. Rev. A {\bf 41}, 3258 (1990). 
\bibitem{avraham} ben-Avraham D., Burschka M. A., and Doering C. R., J. Stat. Phys. {\bf 60}, 695 (1990).
\bibitem{alcaraz} Alcaraz F. C.,  Droz M., Henkel  M. and Rittenberg V. , Ann. Phys. {\bf 230}, 250 (1994).
\bibitem{krebs}  Krebs K., Pfannmuller M. P., Wehefritz  B. and Hinrinchsen H. , J. Stat. Phys. {\bf 78}, 1429 (1995).
\bibitem{santos}  Santos J. E., Schutz G. M. and Stinchcombe R. B., J. Chem. Phys. {\bf 105}, 2399 (1996).
\bibitem{schutz}Schutz G. M., Z. Phys. B {\bf 104}, 583 (1997).
\bibitem{oliveira}  de Oliveira M. J., Brazilian Journal of Physics, {\bf 30} 128 (2000). 
\bibitem{kang} Kang K.  and Redner S., Phys. Rev. A {\bf 30}, 2833 (1984); {\bf 32}, 435 (1985).
\bibitem{peliti} Peliti L., J. Phys. A {\bf 19}, L365 (1986).
\bibitem{zumofen} Zumofen G., Blumen A. and Klafter J., J. Chem. Phys. {\bf 82}, 3198 (1985). 
\bibitem{droz} Droz M. and Sasvari L., Phys. Rev. E {\bf 48}, R2343 (1993).
\bibitem{krap_ballistic} Krapivsky P. L. and Ben-Naim E., Phys. Rev. E {\bf 56}, 3788 (1997).
\bibitem{bennaim} Ben-Naim E., Redner S.  and Leyvraz F., Phys. Rev. Lett. {\bf 70}, 1890 (1993). 
\bibitem{krap_ballisticanni_2001} Krapivsky L. and Sire C., Phys. Rev. Lett. {\bf 86}, 2494 (2001).
\bibitem{albano_levy} Albano E. V., J. Phys. A: Math. Gen. {\bf 24}, 3351 (1991). 
\bibitem {krap_bennaim2016} Ben-Naim E., Krapivsky P. L. and Randon-Furling J., J. Phys. A: Math. Theor. {\bf 49}, 205003 (2016).

\bibitem {soham_ray_ps2011} Biswas S., Sen P. and  Ray P., J. Phys.: Conf. Ser. {\bf 297}, 012003 (2011).
\bibitem {ray_ps2015} Sen P. and Ray P., Phys. Rev. E {\bf 92}, 012109 (2015).
\bibitem{ray_daga2019} Daga B. and Ray P., Phys. Rev. E {\bf 99}, 032104 (2019).
\bibitem{pratik_ps2019}Mullick P. and Sen P., Phys. Rev. E {\bf 99}, 052123 (2019).

  
\bibitem{soham_ps2009} Biswas S. and Sen P. , Phys. Rev. E {\bf 80}, 027101 (2009).

 

\end{thebibliography}
\end{document}